\documentclass[twoside,fleqn]{ActaStyle}
\usepackage{times,cite}

\usepackage{lscape}             % if you need landscape enviroinment
\usepackage{graphicx,epsfig}    % if you need to include figures in PostScript
\usepackage[tbtags]{amsmath}    % if you need more math

\def\authorheading{P. Klaja and P. Moskal for the COSY-11 collaboration}

\def\shorttitle{Correlation femtoscopy for studying $\eta$ meson production mechanism}

\begin{document}
%% ---------------------------------------------------

\begin{center}
%%%  page range, first and last page
\pagerange{1}{9}

%%% paper title
\title{%
CORRELATION FEMTOSCOPY FOR STUDYING $\eta$ MESON PRODUCTION MECHANISM.\\
}

P.~Klaja\email{klajus@poczta.onet.pl}$^{,a}$, P.~Moskal$^{a,b}$, H.-H.~Adam$^c$,
A.~Budzanowski$^{d}$, E.~Czerwi{\'n}ski$^a$, 
                      R.~Czy{\.z}ykiewicz$^{a,b}$, D.~Gil$^a$, 
D.~Grzonka$^b$, M.~Janusz$^a$,
                      L.~Jarczyk$^a$, B.~Kamys$^a$,
A.~Khoukaz$^c$,
                      K.~Kilian$^b$, J.~Majewski$^a$, W.~Migda{\l}$^a$, 
                      W.~Oelert$^b$, C.~Piskor--Ignatowicz$^a$,
J.~Przerwa$^a$, J.~Ritman$^b$,
                      T.~Ro{\.z}ek$^{b,e}$,
R.~Santo$^c$, T.~Sefzick$^b$,
                      M.~Siemaszko$^e$, J.~Smyrski$^a$,
                      A.~T{\"a}schner$^c$, 
P.~Winter$^b$, M.~Wolke$^b$,
                      P.~W{\"u}stner$^{f}$,
Z.~Zhang$^b$, W.~Zipper$^e$\\
\vspace{0.5cm}
{\small
%$^a$ M.~Smoluchowski Institute of Physics, Jagellonian University,
%Poland\\
%$^b$ Institut f{\"u}r Kernphysik, Forschungszentrum J{\"u}lich,
%Germany\\
%$^c$ Institut f{\"u}r Kernphysik, Universit{\"a}t M{\"u}nster,
%Germany\\
%$^d$ Institute of Nuclear Physics, Cracow, Poland\\
%$^e$ Institute of Physics, University of Silesia, Katowice, Poland\\
%$^{f}$ ZEL Forschungszentrum J{\"u}lich, Germany}\\
$^a$Institute of Physics, Jagellonian University, 30-059 Cracow, Poland \\
$^b$IKP, Forschungszentrum J\"ulich, 52425 J\"ulich, Germany \\
$^c$IKP, Westf\"alische Wilhelms-Universit\"at, 48149 M\"unster, Germany \\
$^d$Institute of Nuclear Physics, 31-342 Cracow, Poland \\
$^e$Institute of Physics, University of Silesia, 40-007 Katowice, Poland \\
$^f$ZEL, Forschungszentrum J\"ulich, 52425 J\"ulich, Germany} \\

\end{center}
%%% Date of submition
\day{December 12, 2005}

%%% abstract of the paper
\abstract{The high statistics data from the $pp \to pp\eta$ reaction measurement, 
delivered by the COSY-11 collaboration, are now being evaluated 
using the correlation femtoscopy technique. This method is based on the relative 
momentum correlations of two emitted protons and may permit determination 
of the size of the reaction volume.\\ 
For the very first time, we apply an intensity
interferometry technique to study the mechanism of the meson production 
via the nucleon-nucleon interaction close to the kinematical threshold. 
We invented a method to determine
correlation function for the $pp\eta$ system free from the physical multi-pion 
production background. We show the comparison of experimental results 
with theoretical predictions and appraise the accuracy achieved for the 
determination of the size of the emission source. 
}

%%% PASC numbers of your article

\section{Introduction}
\label{sec:intro}
One of the possible mechanisms of the
$\eta$ meson production via the $pp \to pp\eta$ reaction is
a direct emission of the $\eta$ meson from the interaction region and the
other possibility, believed to be dominant, is the creation
of this meson via the resonant state $S_{11}(1535)$ \cite{review, faldt, hanhart}.
Very close to the kinematical threshold the direct influence of the resonant state on 
the form of differential distribution of the cross section (e.g. Dalitz plot) 
is very difficult to observe due to the fact that the resonance is more than an order 
of magnitude broader ($\Gamma$ $\approx$ $150$ MeV \cite{gamma}) than the available range of the 
invariant mass distribution, which in the case considered in this article amounts to 
$15.5$ MeV \cite{prc69}. Therefore it is worth trying to find an observable different 
from the distributions of the cross sections which can deepen our understanding of the 
reaction dynamics. We assess that such an additional information about the 
production mechanism may be derived from 
the size of the reaction region from which the particles emerge.\\ 

In comparison to the direct production, the presence of the resonance in the production 
process ($pp \to pN^{*} \to pp\eta$) will increase the effective size of the protons 
separation at the moment when they appear as free (on-shell) particles. 
The information about the effective spatial proton's separation at this moment may be 
derived from the proton-proton correlation function. 
However, the shape of this function depends also on the interaction 
between the emitted particles, and in order to draw unbiased conclusions about 
the spatial size of the emission source, the effect from the final state
interaction (FSI) has to be determined precisely. 
Due to the presence of the Coulomb and strong interaction, and the non-trival 
problems with the description of the three body (nucleon-nucleon-meson) system, 
there exists at present no rigorous description of the interaction 
within the $pp\eta$ system \cite{kleefeld}. 
Yet, in practice we can account for the $pp\eta$ FSI, 
parametrizing it from the 
precisely measured $m_{p\eta}$ and $m_{pp}$ invariant mass distributions \cite{habil}.
Therefore, by the analogy to the successful analysis
based on the Dalitz plot and correlation function, performed for 
the determination of the size of the neutron halo in the 
$^{14}Be$ ($^{12}Be + n + n$) and the mechanism of its dissociation on C and Pb 
targets \cite{marques}, we would like to conduct  similar investigations for the 
$pp\eta$ system in $pp \to pp\eta$ reaction. 
 In case of the $pp \to pp\eta$ reaction it can be possible
  that two protons coming from the reaction volume could be emitted
  with a uncorrelated phases since the reaction proceeds in two steps
  ($pp\to pN^{*}\to pp\eta$)
  via the excitation of the intermediate resonance state $N^{*}$ (1535)
  which decays delayed in time into the proton and the $\eta$ meson.

Due to the short life-time of the resonant state, the average increase of the 
effective size of the source in respect to the direct production 
is expected to be small (in the order of the fraction of femtometer), but 
as we will argue later, the accuracy we can achieve is better than 
$0.1$ fm. Moreover, there is another interesting aspect connected to the study
of the reaction volume. Namely, it cannot be a priori excluded  that the 
geometrical dimensions of the source can be much larger 
due to the reflection of the topological (Borromean) bounding in the $pp\eta$ 
system. As Borromean we call a bound three-body system in which none of the two-body
subsystems is bound.
There exists, an encouraging and interesting conjecture of 
S.~Wycech who pointed out that a large enhancement in the excitation function of the
$pp \to pp\eta$ reaction observed close to the kinematical
threshold may be described assuming,
that the proton-proton pair is produced from a large (Borromean like) object 
of a $4$~fm radius~\cite{wycech}. \\

The Borromean 
bounding in the three body system is a very intriguing issue attracting 
the researchers from different fields of science.
In nuclear physics the $^{11}Li$ ($^9Li + n + n$) and $^6He$ ($\alpha + n + n$)
nuclei have been found to have such a property \cite{zhukov}. 
 Another, interesting and worth noticing example, 
is that very recently 
nanoscale Borromean rings were constructed in a wholly synthetic
molecular form~\cite{chichak, cantrill}.\\
At present it is, however, still not established whether
the low energy $pp\eta$ system can form a Borromean or resonant state. 

In the subsequent sections we will present a 
correlation function for the $pp\eta$ system produced in the collisions of nucleons, 
where the $\eta$ meson is identified via the missing mass technique. In such measurements 
it is impossible to identify the production of $pp\eta$ system on the event-by-event 
basis due to the physical background originating from the $pp \to pp pions$ reactions. 
We have, however,  succeeded to develop a method for constructing a correlation function 
free from the multi-pion production background. The technique will be described in 
this article.

\section{Correlation femtoscopy}
\label{sec:corr_femt}
The momentum correlations of particles at small relative velocities are 
widely used to study space-time characteristics of production processes, 
so serving as a correlation femtoscopy \cite{nukl}. Correlation femtoscopy 
originates from an intensity interferometry known as HBT effect \cite{hbt}. 
Implemented into nuclear physics \cite{koonin, kopyl, nukl}, this technique permits to 
determine the size of the source from which the particles are emitted.
It is based on the correlation function,
which relates the space-time separation of the particles to 
their momenta $p_{1}$ and $p_{2}$ at the emission time.
This function can be expressed in terms of pair- and single-particle
cross section~\cite{boal}:
\begin{equation}
R(p_{1},p_{2})= C \cdot \frac{d^6\sigma/d^3p_1 d^3p_2}
                             {(d^3\sigma/d^3p_1)(d^3\sigma/d^3p_2)} - 1,
\label{equ:1}
\end{equation}
where C denotes the overall normalization constant. 
Generally one can relate a correlation function with a Fourier transform 
of the spatial distribution of the emission source~\cite{baym}.
The shape of a two-proton correlation
function depends on the spatial size of the source, quantum statistics and 
the interaction between the outgoing particles.  

In our case the influence of the $pp\eta$ FSI can be derived 
from the experiment. More details will be given in the next section.
Thus,  taking FSI into account when calculating a correlation function 
will permit to study the dependence of the shape of this function on
the spatial dimensions of the reaction region. 

As far as the experiment is concerned,
in case of impossibility for the single particle cross section measurement, 
an alternative definition for the correlation function can be applied. 
Equivalently to the equation \ref{equ:1}, one can calculate such a function 
by generating an "uncorrelated" yield via event mixing technique, 
as shown by Kopylov and Podgoretsky \cite{kopyl}. 
The correlation function is then given by \cite{boal}:
\begin{equation}
    R(q)+1 = C^*_{12}~\frac{Y_{12}(q)}{Y^*_{12}(q)},
\label{equ:2}
\end{equation}
where $Y_{12}(q)$ denotes the coincidence yield and $Y^*_{12}(q)$  stands for 
the yield derived from the uncorrelated reference sample. 
$C^*_{12}$ is an appropriate normalization constant. Here, $R(q)$ denotes 
a projection of the correlation function 
onto the relative momentum of emitted
particles $q = |\vec{p_{1}}-\vec{p_{2}}|$.

\section{Interaction within the $pp\eta$ system}
\label{sec:ppeta}
\begin{figure}[t]
\begin{center}
\includegraphics[width=7.0cm]{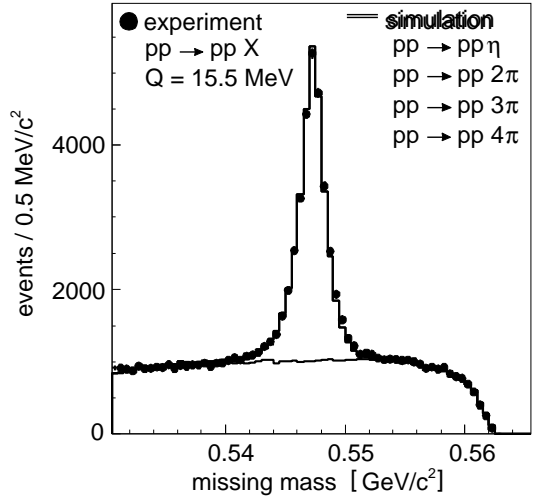}
\end{center}
\caption{
Missing mass spectrum for $pp \to ppX$ rection determined in the experiment at 
beam momentum $p_B = 2.0259$ GeV/c \cite{prc69} (points). 
The solid line histograms present the simulation of $1.5$~$\cdot$ 
$10^{8}$ events of 
$pp \to pp\eta$ reaction, and $10^{10}$ events for the 
$pp \to pp2\pi$, $pp \to pp3\pi$ and $pp \to pp4\pi$ reactions. The simulated 
histograms were fitted to the data varying only the magnitude.
}
\label{fig:missmass}
\end{figure}

\begin{figure}[t]
\parbox{0.5\textwidth}{\psfig{file=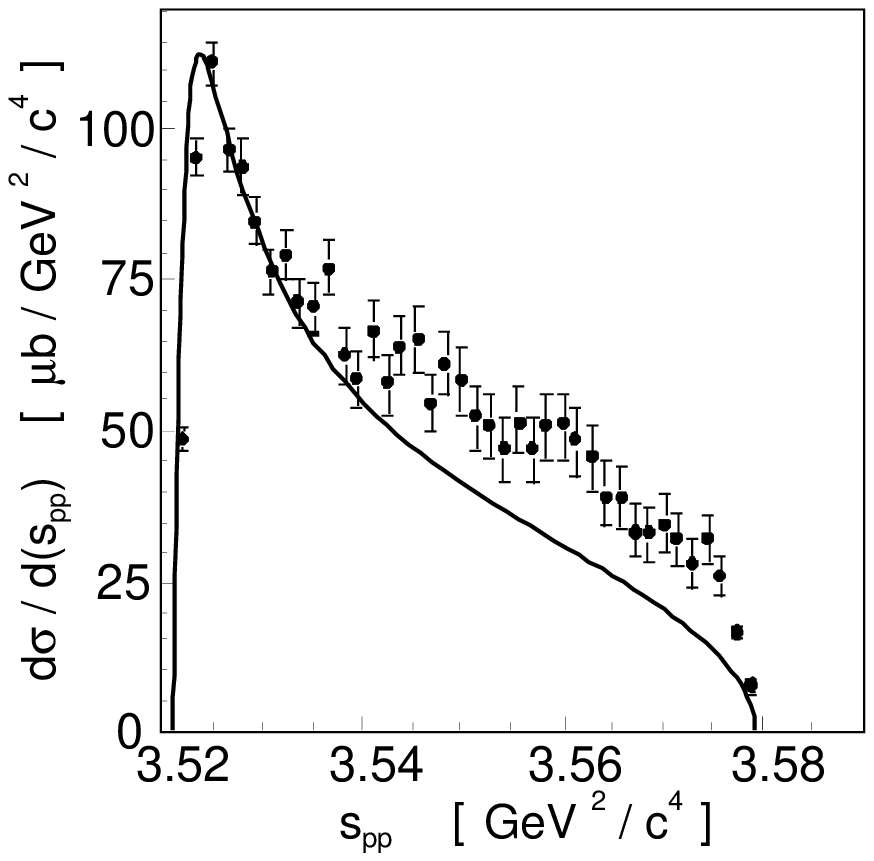,width=5.9cm}}
\parbox{0.5\textwidth}{\psfig{file=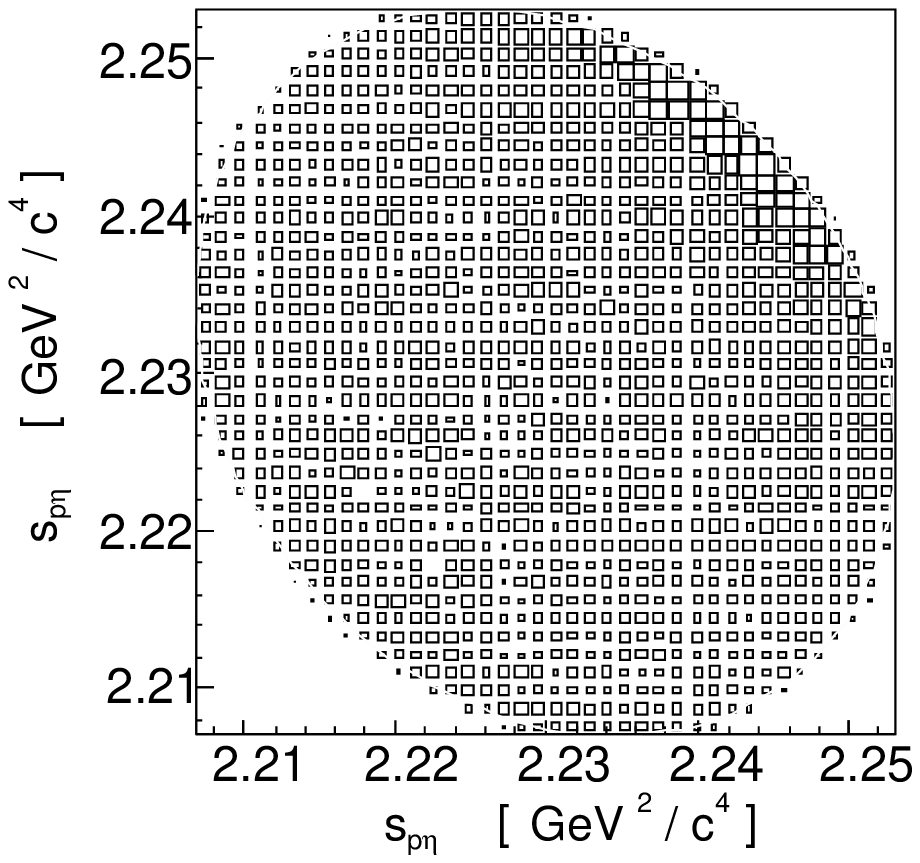,width=6.1cm}}
\caption{
{\bf (Left panel):} Distribution of the square of the proton-proton 
($s_{pp}$) invariant mass for the $pp \to pp\eta$ rection at $Q = 15.5$~MeV 
\cite{prc69, habil}.
The solid line corresponds to the calculations under the assumption 
of only proton-proton interaction~\cite{baru}. 
{\bf (Right panel):} Dalitz plot distribution. Experimental results 
determined for the $pp \to pp\eta$ rection at the excess energy of 
$Q = 15.5$~MeV. Data were corrected for the detection acceptance and efficiency.
%Both figures are adapted from \cite{prc69}.
}
\label{fig:fsi_dalitz}
\end{figure}

The influence of the interaction between ejectiles of the $pp \to pp\eta$ reaction 
on the correlation function can be inferred from the invariant mass distributions 
of the two-particles subsystems.
The $pp \to pp\eta$ reaction was measured with high statistics by the 
COSY-11 collaboration at beam momentum $p_B = 2.0259$~GeV/c \cite{prc69}. 
The determined missing mass spectrum for the $pp \to ppX$ process is presented in 
figure \ref{fig:missmass}.
One can easly recognize the sharp peak from the $\eta$ meson production over a
flat multi-pion production background. The estimated number of $\eta$ production 
events is around 25 thousands. High statistics in an
such experiment allows to derive the distributions of differential cross
sections free of the multi-meson production background \cite{prc69, habil}. 
An example of such a distribution is presented in the left panel of figure \ref{fig:fsi_dalitz}.
This figure presents the projection 
of the phase-space distribution onto the two-proton invariant mass axis. 
One can easily 
recognize the growth of the population density at the range where 
protons have small relative momenta. The shown distribution is free of the 
multi-pion background since the number of $pp \to pp\eta$ events has been 
extracted for each point separately, first selecting data according to the 
square of the proton-proton invariant mass ($s_{pp}$) value and then for each 
$s_{pp}$ interval constructing a missing mass 
distributions. Generally, the full experimentally accessible information of the interaction 
within the $pp\eta$ system is contained in the Dalitz plot distribution (shown in the 
right panel of figure~\ref{fig:fsi_dalitz}) which we can use 
when taking into account an influence of the $pp\eta$ FSI on the correlation function. 

\section{Proton-proton correlation function for the $pp \to pp\eta$ reaction}
\label{sec:exp}
In this section we would like to present a method of 
evaluation of proton-proton correlation function 
for the investigated $pp \to pp\eta$ reaction.
Subtraction of the physical background and corrections for the limited acceptance 
of the detection system constitute 
the two main challanges we have to face when deriving the correlation function
from the experimental data. Here we will concentrate on the former one.

\begin{figure}[h]
\parbox{0.5\textwidth}{\psfig{file=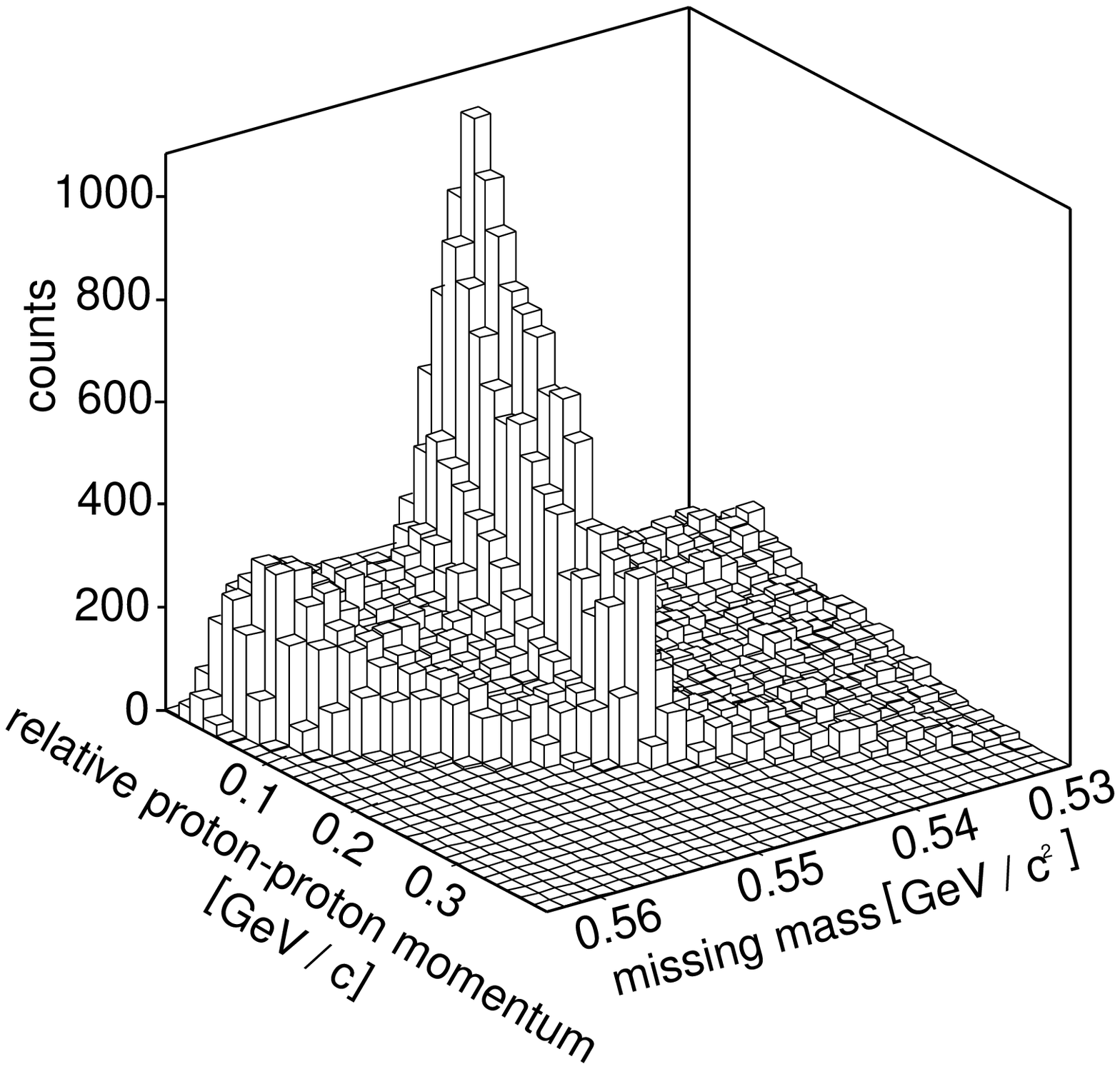,width=6.0cm}}
\parbox{0.5\textwidth}{\psfig{file=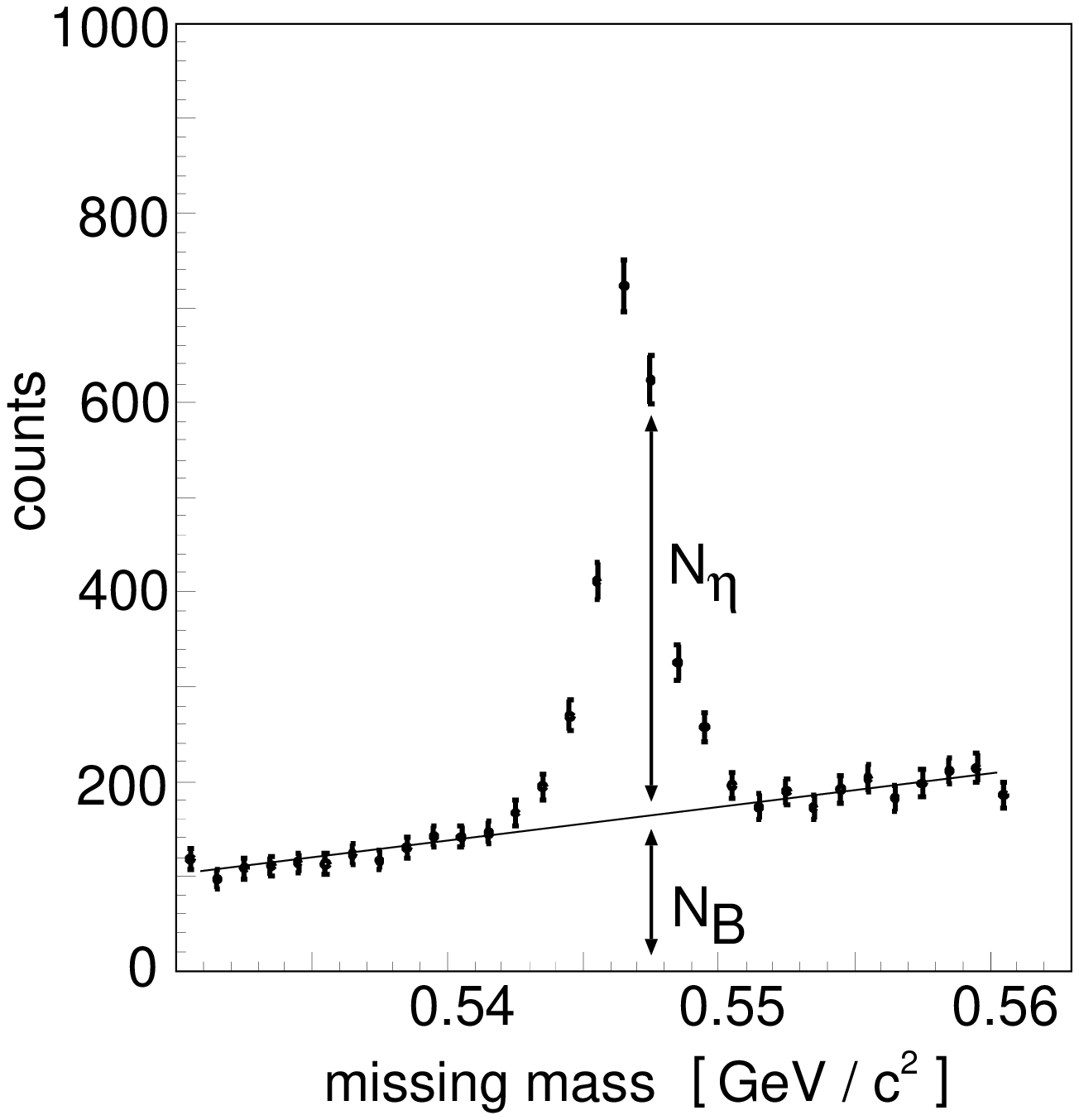,width=6.0cm}}
\caption{
{\bf (Left panel):} The example of the high statistics distribution 
of the measured $pp \to ppX$ reaction 
close to the $\eta$ meson production threshold \cite{prc69}.
{\bf (Right panel):} The example of missing mass spectrum measured \cite{prc69} 
for the $pp \to pp\eta$
reaction at $q~\in~(0.06;~0.07)$ GeV/c.
}
\label{fig:stat_back}
\end{figure}

According to the equation \ref{equ:2}
the two-proton correlation function $R(q)$  can be defined~\cite{boal}
as a ratio of the reaction yield $Y_{pp\eta}(q)$ to the uncorrelated
yield $Y^*(q)$.
In the discussed $pp \to pp\eta$ experiment, only 
four-momenta of two protons were measured and the unobserved meson
was identified via the missing mass technique \cite{prc69, habil}.
In such measurement the entire accessible information about an event is contained
in the momentum vectors of registered protons.
Therefore,  it is in principle impossible
to decide whether a given event corresponds to the $\eta$ meson production
or whether it is due to the  multi-pion creaction. 
However, statistically, on the basis of the missing mass spectra,
one can derive a number of events originated from the production of $pp\eta$ system,
for a chosen region of the phase-space.
Therefore, $Y_{pp\eta}(q)$ can be easily extracted for each studied interval 
of $q$ by  dividing the sample of measured events according
to the value of $q$, next calculating 
the missing mass spectra of the $pp \to ppX$ reaction for each sub-sample separately,
and counting the number of $pp\eta$ events from these spectra.
An example of such histogram for one value of $q$ is presented 
in the right panel of figure~\ref{fig:stat_back}. To demonstrate the full set 
of the data in the left panel of figure~\ref{fig:stat_back} we show also
a spectrum of the population density over the plane 
spanned by $q$ and the missing mass.

The statistics obtained in the considered measurement allowed to divide  
the kinematically available range 
of $q$ into bins whose width ($\Delta{q}~=~5$~MeV/c)
corresponds approximately to the  accuracy of the 
determination of the relative proton momentum~($\sigma(q)~\approx~6$~MeV/c). 

An extraction of $Y^*(q)$, unbiased by the multi-pion production is however not trivial. 
Applying a 
mixing technique one can   
construct an uncorrelated reference sample
taking momentum vectors of protons corresponding  to different real events.
A real event is determined by the  momentum vectors of two protons registered in coincidence, 
and an uncorrelated event will thus comprise  momentum vectors of protons ejected from different reactions.
Unfortunately, in such a sample of uncorrelated momentum vectors,
due to the loss of the kinematical bounds,
the production of the $\eta$ meson
will be not reflected on the missing 
mass spectrum and hence it can not be used to extract a 
number of mixed-events corresponding to the production of the $\eta$ meson.
Therefore, in order to determine a background-free correlation function
for the $pp \to pp\eta$ reaction we performed an analysis in the following manner:
First, for each event,  we have determined  a probability $\omega$
that this event corresponds to the $pp \to pp\eta$ reaction.
The probability $\omega_{i}$, that $i^{th}~pp \to ppX$ event with a missing 
mass $m_{i}$, and relative momentum of $q_{i}$ corresponds to $pp \to pp\eta$
reaction was estimated according to the below formula:
\begin{equation}
    \omega_{i} = \frac{N_{\eta}}{N_{\eta} + N_{B}}(m_{i},q_{i}),
\label{equ:4}
\end{equation}
where $N_{\eta}$ stands for the number of 
the $pp \to pp\eta$ reactions and $N_{B}$ is the number of events corresponding to the multi-pion
production. The values of $N_{\eta}$ and $N_{B}$ 
were extracted from the missing mass distributions produced 
separately for each of the studied intervals of
relative momentum $q$. An example of a missing mass spectrum 
with the pictorial definitions of $N_{B}$ and $N_{\eta}$ is presented in the right panel of figure~\ref{fig:stat_back}.
Next, in order to obtain a value of $Y_{pp\eta}(q)$ for a certain $q$
we have added probabilities of all events for which a relative momentum of two 
 protons
belongs to the bin centered at $q$.
Thus, 
$Y_{pp\eta}(q) = \sum_i{\omega_i}$ where $i$ enumerates all events with relative protons momentum 
corresponding to the considered $q$ bin.
The result is of course per definition the same as obtained earlier by counting the number
of events ($Y_{pp\eta}(q)~=~\sum_j{N_{\eta}(j)}$) 
where $j$ enumerates bins in the histogram.

However, now having introduced the weights we can calculate  also the value of $Y^*(q)$ 
without a bias of the multi-pion background.
We can achieve this by sorting an uncorrelated sample according to the $q$ values similarly as
in the case of the correlated events and next for each sub-sample 
we construct background free $Y^*(q)$ as a sum of the probabilities that both protons in
an uncorrelated event originate from the reaction where the $\eta$ meson was created.
Specifically,  if in a given uncorrelated event denoted by $k$, one momentum is taken 
from a real event say $k1$
and the second momentum from an real event $k2$, then the probability that both correspond 
to the reactions where the $\eta$ was created equals to $\omega_{k1} \cdot \omega_{k2}$,
and hence the uncorrelated yield $Y^*(q)$ may be constructed as a $\sum_k{\omega_{k1} \cdot \omega_{k2} }$,
where $k$ enumerates events in the uncorrelated sub-sample with a relative protons momentum $q$.

In the limit of absence of background all values of $\omega$ would be equal to 1 and so deriving 
both $Y_{pp\eta}(q)$ and $Y^*(q)$ we would just count
the events. 
In the presence of the background if a missing mass of real event is from the outside of the
$\eta$ signal than it is certain that this event corresponds to the multi-pion production process
and it indeed will not be taken into account in the calculation of $Y_{pp\eta}(q)$ and $Y^*(q)$
since its  weight  will be equal to zero.
If the missing mass of the given event is from the region where $N_{\eta}$ is larger than zero
than we don't know  whether the event is from the $\eta$ or multi-pion creation. However, 
for the calculation of the yields we need  only the overall number of $\eta$ events from 
a given range of the $q$ value.  This could be determined either by taking  only 
$\eta$ events into account and counting them with a weight equal to one (which is in principle impossible
since this events are kinematically indistinguishable from the multi-pion production)
or as we did by counting all events fulfilling a required kinematical conditions (in our case we look only at $q$)
and assigning them  probabilities that they belong to the $\eta$ production.

\begin{figure}[t]
\begin{center}
\includegraphics[width=7.0cm]{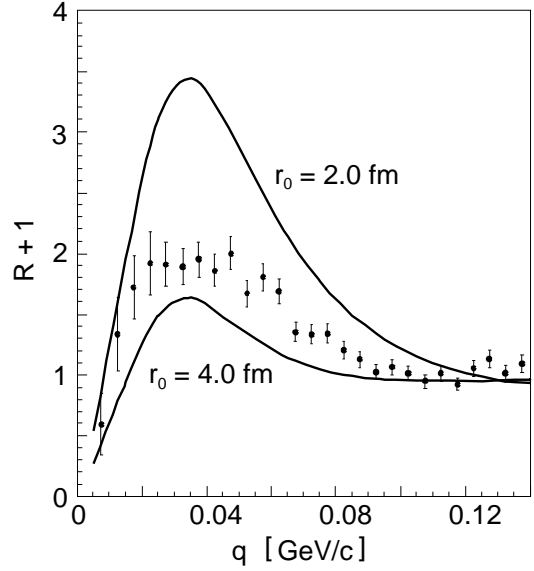} 
\end{center}
\caption{
Comparison of the experimental correlation function for the 
$pp \to ppX$ reaction, without acceptance correction represented by 
full dots and theoretical calculations indicated as two solid lines 
for emission sources parametrized by Gaussian distribution with 
$r_{0}~=~2.0$~and~$4.0$~fm. 
}
\label{fig:results}
\end{figure}
The correlation function derived from the data is presented in
figure~\ref{fig:results}. 
It is compared to the 
calculations, performed assuming a simultaneous emission of the two protons
and the $\eta$ meson and approximating tentatively the effective spatial shape
of the emission zone by the Gaussian distribution. 
In such a case the standard deviation of such distribution -~hereafter
referred to as $r_{0}$~-
constitutes the measure of the dimension of the source. 
For the simulations we presently adapted a computing procedure written by
R. Lednicky~\cite{lednicky, lyubo}, which already has been successfully applied
in other studies (example in ref. \cite{marques}). 
At present we take into account proton-proton
FSI only, but in further analysis, we would like to 
make simulations with compliance of the $pp\eta$ FSI as can be  
extracted from the Dalitz plot distributions.

The determined experimental correlation function 
shows a maximum at a value of $q$ as predicted by simulations.
The height of the peak at q~$\approx~40$~MeV/c
depends significantly on the value of $r_0$ and therefore the
magnitude of this maximum may serve
as a measure of the volume of the reaction zone. 
A rough comparison between theoretical correlation functions simulated with 
$r_{0} =~2$~fm and $4$~fm, respectively, 
and the experimental one shown in the figure~\ref{fig:results} indicates
that the size of the reaction volume
can be approximated by the Gaussian distribution with $r_0~\approx 3$~fm.
Of course, generally
the size of the reaction zone can be extracted by the comparison of the
experimental and simulated correlation functions treating $r_0$ as a
fitting parameter.

The results presented in the figure \ref{fig:results} have not been corrected for the 
acceptance of the COSY-11 detection setup yet,  and the theoretical calculations
have been performed without taking into account the experimental spread of the momenta.
These corrections could significantly 
influence the final results.
Therefore in a present article we do not perform a quantitative analysis. 
We only put out the emphasis on exploring the method which can be further adapted to 
examine the dynamics of $\eta$ and $\eta'$ meson production mechanism in nucleon-nucleon 
collisions. 
Although after the mentioned corrections  the result may be different,
       it is worth noticing that the extracted value of $r_0$ is compatible with the
       effective range (2.8~fm~\cite{naisse}) of the proton-proton interaction.

From the comparison of the theoretical lines and the data in figure~\ref{fig:results}
one can appraise a statistical accuracy of the $r_0$ determination which can be achieved
when performing fit with $r_0$ as a free parameter.
Examining in figure~\ref{fig:results} the range of $0.02$~GeV/c~$< q <~0.10$~GeV/c
one observes there $15$ points with a statistical errors corresponding to 
the accuracy varying between $0.2$~fm and $0.5$~fm. Thus as far as the statistical 
precision is concerned we will be able to determine the effective size of 
the source with the precision much better than the errors quoted above for 
the single points. 

\begin{ack}
We acknowledge the support of the
European Community-Research Infrastructure Activity
under the FP6 "Structuring the European Research Area" programme
(HadronPhysics, contract number RII3-CT-2004-506078),
of the FFE grants (41266606 and 41266654) from the Research Centre J{\"u}lich,
of the DAAD Exchange Programme (PPP-Polen),
of the Polish State Committe for Scientific Research
(grant No. PB1060/P03/2004/26), \\
and of the
RII3/CT/2004/506078 - Hadron Physics-Activity -N4:EtaMesonNet.
\end{ack}


\newpage
\begin{thebibliography}{22}

\bibitem{review} 
P. Moskal, M. Wolke, A. Khoukaz, W. Oelert: \emph{Prog. Part. Nucl. Phys.} {\bf 49} (2002) 1.
%%CITATION = HEP-PH 0208002;%%

\bibitem{faldt} 
G. F{\"a}ldt, T. Johansson, C. Wilkin: \emph{Phys. Scripta} {\bf T~99} (2002) 146.
%%CITATION = PHSTB,T99,146;%%

\bibitem{hanhart} 
C. Hanhart: \emph{Phys. Rept.} {\bf 397} (2004) 155.
%%CITATION = HEP-PH 0311341;%%

\bibitem{gamma} 
S. Eidelman et al.: \emph{Phys. Lett.} {\bf B~592} (2004) 1.

\bibitem{prc69} 
P. Moskal et al.: \emph{Phys. Rev.} {\bf C~69} (2004) 025203.
%%CITATION = NUCL-EX 0307005;%%

\bibitem{kleefeld} 
F. Kleefeld: Schriften des FZ-J{\"u}lich, Matter
                   and Material 11 (2002) 51,\\
		   e-Print Archive: nucl-th/0108064;
                   and this proceedings, e-Print Archive: nucl-th/0510017. 
		   
\bibitem{habil} 
P. Moskal: e-Print Archive: hep-ph/0408162.

\bibitem{marques} 
F. M. Marqu{\'e}s et al.: \emph{Phys. Rev.} {\bf C~64} (2001) 061301(R).
%%CITATION = NUCL-EX 0101004;%%

\bibitem{wycech} 
S. Wycech: \emph{Acta Phys. Pol.} {\bf B~27} (1996) 2981.
%%CITATION = APPOA,B27,2981;%%

\bibitem{zhukov} 
M. V. Zhukov et al.: \emph{Phys. Rept.} {\bf 231} (1993) 151.
%%CITATION = PRPLC,231,151;%%

\bibitem{chichak} 
K. S. Chichak et al.: Science {\bf 304} (2004) 1308.

\bibitem{cantrill} 
S. J. Cantrill et al.: \emph{Acc. Chem. Res.} {\bf 38} (2005) 1.

\bibitem{nukl} 
R. Lednicky: Nukleonika {\bf 49~(Sup.~2)} (2004) S3.

\bibitem{hbt} 
R. Hanbury-Brown, R. G. Twiss: \emph{Phil. Mag.} {\bf 45} (1954) 663.
%%CITATION = PHMAA,45,663;%%

\bibitem{koonin} 
S. E. Koonin: \emph{Phys. Lett.} {\bf B~70} (1977) 43.
%%CITATION = PHLTA,B70,43;%%

\bibitem{kopyl} 
G. I. Kopylov, M. I. Podgoretsky: \emph{Sov. J. Nucl. Phys.} {\bf 15} (1972) 219.
%%CITATION = SJNCA,15,219;%%

\bibitem{boal} 
D. H. Boal et al.: \emph{Rev. Mod. Phys.} {\bf 62} (1990) 553.
%%CITATION = RMPHA,62,553;%%

\bibitem{baym} 
G. Baym: \emph{Acta Phys. Pol.} {\bf B~29} (1998) 1839.
%%CITATION = NUCL-TH 9804026;%%
 
\bibitem{baru} 
V. Baru et al.: \emph{Phys. Rev.} {\bf C~67} (2003) 024002.
%%CITATION = NUCL-TH 0212014;%%

\bibitem{lednicky} 
R. Lednicky: private communication (2005).

\bibitem{lyubo} 
R. Lednicky and L. Lyuboshits: \emph{Sov. J. Nucl. Phys.} {\bf 35} (1982) 770.
%%CITATION = SJNCA,35,770;%%

\bibitem{naisse} 
J. P. Naisse: \emph{Nucl. Phys.} {\bf A~278} (1977) 506. 
\end{thebibliography}
\end{document}